# Genotype networks, innovation, and robustness in sulfur metabolism.

João F. Matias Rodrigues[1,3] and Andreas Wagner [1,2,3]

[1] *Department of Biochemistry Bldg. Y27, University of Zurich, Winterthurerstrasse 190, CH-8057 Zurich, Switzerland*

[2] *The Santa Fe Institute, 1399 Hyde Park Road, Santa Fe, New Mexico 87505, United States of America*

[3] *The Swiss Institute of Bioinformatics,* Quartier Sorge*, CH-1015 Lausanne-Dorigny Switzerland*

Email addresses:

JFMR: j.rodrigues@bioc.uzh.ch

AW: aw@bioc.uzh.ch

## Abstract

### *Background*

Metabolic networks are complex systems that comprise hundreds of chemical reactions which synthesize biomass molecules from chemicals in an organism's environment. The metabolic network of any one organism is encoded by a metabolic genotype, defined by a set of enzyme-coding genes whose products catalyze the network's reactions. Each metabolic genotype has a metabolic phenotype, such as the ability to synthesize biomass on a spectrum of different sources of chemical elements and energy. We here focus on sulfur metabolism, which is attractive to study the evolution of metabolic networks, because it involves many fewer reactions than carbon metabolism. It is thus more tractable to computational methods that predict metabolic phenotypes from genotypes. Specifically, we study properties of the space of all possible metabolic genotypes, and analyze properties of random metabolic genotypes that are viable on different numbers of sulfur sources.

### *Results*

We show that metabolic genotypes with the same phenotype form large connected genotype networks -- networks of metabolic networks -- that extend far through metabolic genotype space. How far they reach through this space is a linear function of the number of super-essential reactions in such networks, the number of reactions that occur in all networks with the same phenotype. We show that different neighborhoods of any genotype network harbor very different novel phenotypes, metabolic innovations that can sustain life on novel sulfur sources. We also analyze the ability of evolving populations of metabolic networks to explore novel metabolic phenotypes. This ability is facilitated by the existence of genotype networks, because different neighborhoods of these networks contain very different novel phenotypes.

### *Conclusions*

In contrast to macromolecules, where phenotypic robustness may facilitate phenotypic innovation, we show that here the ability to access novel phenotypes does not monotonically increase with robustness.

## Background

In any biological system, genotypes contain the information needed to make phenotypes. The relationship between genotype and phenotype is also known as a genotype-phenotype map [1]. The ability to analyze different kinds of biological systems computationally has allowed a detailed characterization of genotype-phenotype maps for different systems. One common feature of genotype-phenotype maps is the existence of genotype networks, connected sets of genotypes that adopt the same phenotype. They exist in systems as different as model proteins [2], RNA secondary structures [3], regulatory circuits [4], and metabolic networks [5, 6]. Another feature is the large phenotypic diversity that is found in different neighborhoods of a genotype network [3, 4, 5, 6]. These two properties facilitate the exploration of novel and potentially beneficial phenotypes in genotype space. By analyzing genotype-phenotype maps of different systems, one can identify general features of genotype maps, as well as features that are specific to a system.

In this work we concentrate on the genotype-phenotype maps of metabolic networks involved in the utilization of sulfur. Our motivation is twofold. First, studying sulfur metabolism allows us to examine the generality of earlier observations made for the genotype-phenotype map of carbon metabolism [5, 6]. It gives us insight into properties of metabolic genotype-phenotype maps that are not restricted to just one class of metabolic systems. Second, it allows us to characterize the organization of metabolic genotype space, and to study evolving *populations* of networks in this space. Carbon metabolism comprises so many reactions that the computational demands of studying population processes in its genotype space are too high for current computational technology. Sulfur metabolism, in contrast, comprises a smaller number of chemical reactions, which renders the computational analysis of population processes more tractable.

Despite being involved in fewer reactions, sulfur is no less essential to biological organisms than other elements, such as carbon or nitrogen. Sulfur is a versatile and integral element in the biochemistry of organisms [7, 8]. Its presence in biological organisms ranges from 0.5% to 50% of dry weight [7]. It occurs in multiple oxidation states, ranging from the highly oxidized $S^{4+}$ to the reduced state $S^{2-}$. This versatility in oxidation state may explain the diversity of sulfur metabolism and why it is involved in both anabolism as well as catabolism. In catabolism, depending on the environment, sulfur can be used as an electron acceptor or an electron donor, and in some cases even both as donor and acceptor. In anabolism, sulfur must first be reduced in a sequence of energetically expensive steps before being incorporated into biomass [7].

Sulfur is present in two major constituents of biomass, the amino-acid cysteine, which confers stability to proteins through disulfide bonds, and the amino acid methionine, which is the first amino acid of many proteins. Sulfur is also a part of S-adenosylmethionine (also known as AdoMet or SAM). This compound is a cysteine metabolite that is a major methyl donor to the methyl carrier metabolite tetrahydrofolate, which is indispensable for amino acid synthesis, and for the methylation of biomolecules. Furthermore, sulfur is the active element in coenzyme-A, an acyl carrier metabolite involved in the calvin cycle and in lipid synthesis. Sulfur is also present in the active core of iron-sulfur proteins, which are involved in a number of important reactions. Examples include nitrogenase, which enables the fixation of nitrogen, and hemoglobin, which enables the transport of oxygen. Another prominent molecule involving sulfur is glutathione, a peptide responsible for protection against oxidative stress in cells.

In this work, we explore the genotype space of metabolic networks that can use different sources of sulfur and incorporate them into a cell's biomass. We first analyze minimal metabolic networks from which no reactions can be removed without destroying their viability. These metabolic networks can be very different from one another but they all share a subset of reactions that we call superessential. We show the existence of long phenotype-preserving paths through metabolic genotypes space that allow exploring this space through many single phenotype-preserving mutations. Such paths exist also for minimal metabolic networks.

Furthermore, we show that the maximum length of these paths and metabolic network size are linked through a linear function of the number of superessential reactions. Next, we show that the existence of neutral paths allows evolving metabolic networks to encounter an increasing number of novel phenotypes. We finally explore the relationship between robustness and a population's ability to access novel phenotypes through changes in a network's reactions. In contrast to macromolecules, where robustness may facilitate phenotypic innovation [9, 10], we find that the ability to find novel phenotypes in our system peaks at intermediate robustness.

## Results

### *The model.*

We follow an approach taken in a previous study of large-scale metabolic networks [5]. We define a metabolic *genotype* as the set of biochemical reactions that may take place in an organism, and that are catalyzed by gene-encoded enzymes. The set of all reactions used in this work is a subset of 1221 reactions out of 5871 reactions we curated previously [5] from the Kyoto Encyclopedia of Genes and Genomes (KEGG) [11]. These reactions comprise all elementally-balanced reactions that involve sulfur containing metabolites (see methods for details).

A metabolic genotype can be represented in at least 2 different ways (Figure 1). The first views it as a metabolic network graph whose nodes are metabolites. Reactions are represented as directed links from substrate metabolites to product metabolites (Figure 1A). The second views it as a list of reactions (Figure 1C), or, equivalently a binary vector whose length -- 1221 reactions in our case -- corresponds to the number of reactions in a known reaction "universe". Each position *i* in this vector corresponds to a reaction. Its values ('0') or ('1') at position *i* indicate the inability or ability of the organism to catalyze the corresponding reaction (Figure 1C). We define the *phenotype* of a metabolic network as the subset of sulfur sources (out of 124 possible sources we consider, see Methods) that allow the network (metabolic genotype) to synthesize all biomass components, if one of the sulfur sources is provided as the sole sulfur source to the organism. We represent this phenotype as a binary vector of length 124 whose entry at position *i* indicates viability if sulfur source *i* is the sole sulfur source (Figure 1D).

To determine metabolic phenotypes from genotypes, we use flux balance analysis [12], a computational method to find a growth-maximizing steady-state metabolic flux through all reactions in a metabolic network. This method requires information about the stoichiometry of every metabolic reaction, a maximally allowed flux of each metabolite in and out of the environment, and information about an organism's biomass composition (see Methods for details). We focus on a metabolic network's qualitative ability to produce all sulfur-containing biomass precursors. We will study networks that are able to do so from each one of a specific set of sole sulfur sources. For brevity, we call such networks *viable*. We will also refer to the number *S* of sulfur sources that a metabolic network must be viable on as the *environmental demand* imposed on the network.

We next introduce the concept of a genotype network for metabolic networks (Figure 1B) [5]. The nodes in this network correspond to individual genotypes (metabolic networks) *with the same phenotype*. Two genotypes are linked – they are *neighbors* -- if they differ in a single reaction. A genotype network thus is a network of metabolic networks. This concept is useful when we examine the evolution of metabolic networks through the addition and elimination of metabolic reactions, which can occur, for example, by horizontal gene transfer [13, 14], or through loss-of-function mutations in enzyme-coding genes. Consider the metabolic network genotype $G_1$ of some organism. This genotype is a node on the genotype network associated with this genotype's phenotype. If some variant $G_2$ of this network -- obtained through an addition or a deletion of a reaction -- has the same phenotype as $G_1$, it will be a neighbor of $G_1$ on the same genotype network. In this manner, one can envision phenotype-preserving evolutionary change of metabolic genotypes as a path through a genotype network. Such paths correspond to successive hops from genotype to genotype, by way of the edges connecting neighboring genotypes (Figure 1B). For our analysis, it will be useful to define a *distance D* between two metabolic network genotypes as the fraction of reactions in which two metabolic networks differ, or

$$D = 1 - \frac{2R_c}{N_1 + N_2},$$

where $R_c$ is the number of reactions shared by both networks and $N_1$ and $N_2$ are the sizes of the compared metabolic networks.

## *Minimal viable metabolic networks can be diverse and contain many superessential reactions.*

We begin with an analysis of minimal viable networks, which provides insights into the reactions that are essential to utilize a specific set of sulfur sources. We define a minimal metabolic network as a network from which not a single metabolic reaction can be eliminated without rendering it inviable. For any one given phenotype *P*, there may be multiple viable minimal networks. We can generate random minimal networks by starting from a network comprised of all 1221 reactions -- which is viable on all sulfur sources -- and eliminating randomly chosen reactions one-by-one, until the network is no longer viable on the sulfur sources defined by *P*. We note that a minimal network is not the same as the network with the smallest possible number of reactions, which could be very difficult to find in a vast metabolic genotype space.

We generated 1000 random minimal metabolic networks viable on a given number *S* of sulfur sources (see methods). Specifically, we generated 100 minimal networks for 10 random sets of sulfur sources with the same number *S* -- but not necessarily identity -- of sources. We note in passing that such networks often also happen to be viable on additional, not required sulfur sources (Figure S1). Figure 2A shows the distribution of genotype distances for pairs of minimal metabolic networks viable on *S*=1, 20, or 60 sulfur sources. The figure demonstrates that, first, random minimal metabolic networks can be very different from one another. Their genotype distance may exceed *D*=0.8, meaning that they may share fewer than 20 percent of reactions. Second, the average distance depends on the number of sulfur sources a network needs to be viable on. Specifically, the average genotype distance is largest ($D_{avg}$=0.6) for minimal metabolic networks viable on *S*=1 sulfur source, and decreases to $D_{avg}$=0.3 for networks viable on *S*=60 sulfur sources. Third, the distribution in genotype distances is much wider for metabolic networks subject to few environmental demands (*S*=1) where it ranges from *D*=0.2 to *D*=0.8, than for metabolic networks subject to many environmental demands (*S*=60) where it ranges from *D*=0.2 to *D*=0.4.

Figure 2B (filled circles) shows the average size of minimal networks as a function of the number of sulfur sources they are viable on. It ranges from 14 reactions for *S*=1 to 87 reactions for *S*=60. By definition, all reactions in a minimal network are essential, but some of these reactions are special. These are reactions that occur in all minimal networks viable on a given set of sulfur sources. We call the reactions we find in all minimal metabolic networks viable on a specific set of sulfur sources *superessential* reactions [6]. The open circles in Figure 2B shows this number of superessential reactions as a function of the environmental demands *S* on a network. The number of superessential reactions increases with *S*, but it is generally much lower than the total number of reactions. For example, at *S*=1, 4 out of 14 reactions are superessential. At *S*=60, 44 out of 87 reactions are superessential. The number of superessential reactions will play an important role in one of our analyses below.

### *Many viable sulfur metabolic network genotypes are connected via paths that lead far through metabolic genotype space.*

We next extended our previous work on carbon metabolism to ask about the existence of genotype networks in the space of sulfur-involving reactions, and of neutral paths that traverse such networks while preserving a metabolic phenotype. A neutral path is a series of mutations (reaction additions or deletions) that leave a phenotype intact (Figure 1B). We emphasize that we do not use the term neutrality in its meaning of unchanged fitness in the field of molecular evolution [15], but merely in the sense of preserving viability on a specific set of sulfur sources.

We were especially interested in two questions. How far does a neutral path typically lead through genotype space? And how does this distance depend on the number *N* of reactions in a network, and on the environmental demands on the network? To answer these questions, we performed 200 random walks of 10’000 mutations each for metabolic networks of various sizes, and for various environmental demands. Specifically, for networks of each size we performed 20 random walks for each of 10 different sets of *S* of sole sulfur sources that we required the network to be viable on. Each random walk started from a random initial viable metabolic network comprising *N* reactions (see methods for details). We allowed *N* to vary by no more than one reaction during the random walk. Moreover, each step in the random walk had to preserve viability. Finally, none of the steps was allowed to decrease the distance to the starting network, in order to maximize the distance from this network (see methods for details).

Figure 3A shows the maximum genotype distance obtained in such random walks for networks up to 300 reactions, where we required viability on $S$=1, 5, 10, 20, 40, or 60 different sole sulfur sources. This distance is in general large. For example, *D* is greater than 0.7 for all metabolic networks with more than 200 reactions. For each value of *S*, the data point at the smallest value of *N* (horizontal axis) corresponds to the minimal metabolic networks we discussed earlier. Perhaps surprisingly, these minimal networks can not only be very diverse, as we saw earlier, but neutral paths starting from any one such network can also reach far through genotype space. For example, the maximal length of neutral paths is $D$=0.65 for minimal metabolic networks viable on $S$=1 sulfur source, and still a sizeable $D$=0.38 for metabolic networks viable on *S*=60 sulfur sources. To provide a point of reference, the *E. coli* metabolic network has 142 reactions involving sulfur. Random viable metabolic networks of this size would have maximum genotype distances between *D*=0.60 (for *S*=60) and *D*=0.96 (for *S*=1).

## *Maximal genotype distance and robustness of metabolic networks are well approximated by simple properties of minimal networks.*

As one adds reactions and entire pathways to a minimal metabolic network, previously essential reactions become non-essential, because the added reactions can compensate for them. This means that the network's fraction of non-essential reactions -- its robustness to reaction deletions -- increases. In addition, the maximal possible genotype distance between metabolic networks of the same size also increases (Figure 3A), because more phenotype-preserving changes become possible as more reactions are added. With these considerations in mind, we asked whether the maximal genotype distance of networks at a given size, as well as their robustness to reaction removal can be predicted from properties of the underlying minimal networks. The answer is yes.

Figure 3A (solid lines) shows the relationship between the maximal genotype distance $D_{max}$ and metabolic network size $N$ as predicted by the equation

$$D_{\max} = 1 - \frac{N_{su}}{N} \qquad (1)$$

Here, $N_{su}$ is the fraction of reactions that are super-essential for a given environmental demand *S*. We had estimated this fraction in our previous analysis from minimal networks (Figure 2B). The simple relationship of equation (1) fits our numerical data (Figure 3A) remarkably well. Only for networks of the smallest size does it systematically overestimate the maximal genotype distance, and does so by no more than 10% percent. We note that our estimates of maximum genotype distances are only lower bounds, such that this discrepancy may be a result of our limited ability to estimate maximal genotype distances accurately. In sum, a simple, linear function of the number of superessential reactions at any one environmental demand $S$ approximates the maximal genotype distance between networks well.

Next we examined how network robustness depends on the size of metabolic networks. We define such robustness as one minus the fraction $f_{ess}$ of essential reactions. Figure 3B and Figure S2, respectively, show the fraction and number of non-essential reactions as a function of network size, for varying environmental demands $S$ on a network. The number of essential reactions decreases linearly with increasing metabolic network size (Figure S2), whereas the fraction of these reactions decreases less than linearly (Figure 3B). Large metabolic networks with 200 reactions or more have a fraction of essential reactions $f_{ess} < 0.4$ for all values of $S$. For smaller metabolic networks ($N<200$), $f_{ess}$ ranges from $f_{ess}=0.1$ under low environmental demands ($S=1$) to $f_{ess}= 0.8$ under high environmental demands ($S=60$).

The solid lines in Figure 3B show that the relationship between the fraction of essential reactions $f_{ess}$ and metabolic network size $N$ is well approximated by the equation $f_{ess}=(N_{min}-m.N)/N=N_{min}/N - m$, which is plotted there. Here, $N_{min}$ is the average size of minimal metabolic networks (estimated above for given $S$) and $m$ is the rate at which the number of essential reactions decreases with increasing metabolic network size (estimated from data in Figure S2). This relationship means that network robustness is a linear function of the ratio $N_{min}/N$, whose inverse indicates how much larger a given network is than a minimal network for a given $S$, and of the rate at which reaction essentiality declines (robustness increases) with increasing $N$.

## *Phenotype diversity in the neighborhood of metabolic networks changes rapidly with genotype distance.*

Thus far, we have concentrated on the characteristics of individual sets of genotypes viable on a given number $S$ of sulfur sources, and on the genotype networks they form. Long paths through a genotype network can contribute to evolutionary innovation in metabolic phenotypes, if many novel phenotypes can be encountered near such a path. We next asked whether this is the case, and how this number of novel phenotypes depends on environmental demands on a network. We consider a phenotype to be novel if it confers viability in a set of new sulfur sources, in addition to those required by the environmental demands imposed on the metabolic network.

We first introduce the notion of a (1-mutant) neighborhood around a metabolic network genotype, which comprises all networks that differ from the genotype by a single reaction (Figure 1B). Because our genotype space has 1221 metabolic reactions, each metabolic network has 1221 neighbors. Of all these neighbors, some will be inviable in any given environment (these are the mutants that have lost an essential reaction), some will maintain the same phenotype, and some will have a novel phenotype while being viable in this environment. That is, they will have gained viability on a new sulfur source. We focus on the latter class of neighbors in this section.

We asked how different are the novel phenotypes in the neighborhood of two metabolic networks $G$ and $G_k$ on the same genotype network, where $G_k$ is a metabolic network derived from $G$ through $k$ random mutations. That is, we determined the fraction of novel phenotypes that occurred in the neighborhood of only one but not the other network. Below we refer to it as the fraction of novel phenotypes *unique* to one neighborhood. If this fraction is very small even for large $k$, then networks in different regions of a genotype space will have mutational access to similar novel phenotypes. Figure 4 shows that the opposite is the case. We obtained the data shown during phenotype-preserving random walks starting from an initial network, by recording the fraction of novel phenotypes that occur in the neighborhood of the mutated metabolic network, but not of the initial network. Every data point is an average over 20 random walks each for 10 different initial metabolic networks (thus, 200 random walks in total) at every value of $S$. Figure 4 shows that the fraction of unique novel phenotypes reaches high values for modest distance between two metabolic networks -- small compared to the maximum genotype distance -- and does not depend much on the number of sulfur sources *S* on which viability is required. It also does not depend strongly on metabolic network size (results not shown). In sum, the neighborhood of moderately different metabolic networks contains very different novel phenotypes.

## *The ability of metabolic networks to encounter novel phenotypes does not depend monotonically on their phenotypic robustness.*

The question of how robustness relates to evolvability has raised considerable interest in recent years [16, 17]. Macromolecules -- RNA and protein -- whose phenotypes are more robust to mutations can access more novel phenotypes than less robust phenotypes [9, 10]. This holds for both large and small evolving populations of such molecules, at least in the case of RNA [10]. We next asked whether these observations are specific to macromolecules, or whether they would hold more generally, that is, for the genotype-phenotype map of our metabolic networks.

Above we considered the robustness of a metabolic *genotype* as its fraction of non-essential reactions. Analogously, we can define the robustness of a metabolic *phenotype* as the average fraction of non-essential reactions of all networks with this phenotype [10]. We showed that this fraction decreases as networks are required to be viable on more and more sulfur sources (Figure 3B). That is, for networks at any given size, the number *S* of sulfur sources on which they are viable can serve as a proxy for phenotypic robustness. The greater a phenotype's *S* is, the smaller is its robustness.

When analyzing how evolving populations explore a genotype network, we need to distinguish between two different kinds of populations. The first are populations where the product of population size and mutation rate is much smaller than one. For brevity, we refer to such populations here as small populations. The second are populations where this product is much greater than one. We refer to these as large populations.

Small populations are genotypically monomorphic most of the time [15], and effectively explore a genotype network much like a single changing network would, i.e., through a random walk on the genotype network. During such a random walk, the changing network encounters different phenotypes in its neighborhood. We determined the cumulative number of different novel phenotypes found in the neighborhood of a random walker. That is, if a phenotype was encountered twice, either in the same neighborhood, or in a neighborhood encountered during an earlier step, we counted it only once. We did so for networks of varying size $N$ and number of sulfur sources $S$. Specifically, for each $N$ and $S$, we carried out 200 random walks of 10'000 mutations each (20 walks for 10 different sets of sulfur sources at each $S$). Figure 5A shows the resulting data. The cumulative number of novel phenotypes is a unimodal function of $S$, indicating that metabolic networks under few and many environmental demands encounter fewer novel phenotypes than under an intermediate number of environmental demands ($S \approx 20$). The cumulative number of novel phenotypes depends strongly on metabolic network size for $S<20$, where larger metabolic networks encounter more novel phenotypes throughout the random walk. It is not sensitive to $N$ for larger values of $S$.

We next turn to the case of large evolving populations. Such populations are polymorphic most of the time. To model their evolutionary dynamics, one needs to track every individual in the population, unlike in the case of monomorphic populations. We determined the cumulative number of novel phenotypes that are mutationally accessible to a population of metabolic networks evolving on (and restricted to) a specific genotype network. This number can be determined by examining, for each generation, the neighborhood of each individual in the population, and by counting the total number of different novel phenotypes encountered. We simulated populations of 100 individuals evolving for 2000 generations (see Methods for details). Figure S3 shows the average number of cumulative unique novel phenotypes accessible to a population through generation 2000. Each data point represents an average and standard deviation over 200 simulations (20 simulations for 10 random sets of sulfur sources at a given *S*). Qualitatively, the figure resembles our observations for a single random walk (Figure 5A), except that the absolute number of cumulative unique phenotypes encountered is higher in the case of evolving populations.

Taken together, these observations show that the number of novel phenotypes accessible to a population does not increase monotonically with phenotypic robustness. It decreases with increasing robustness (decreasing *S*) for low values of *S*, and it increases with robustness at higher values of *S*.

We next examined two candidate explanations of this pattern. The first is that environmental demands and network size affect how rapidly a population can diversify on its genotype network, and thus also how many novel phenotypes it can access. To find out whether this diversification rate matters, we examined the average pairwise genotype distance of our evolving populations. The smaller this difference is, the more slowly a population diversifies. Figure 5A shows a plot of pairwise genotype distances, averaged over an entire population, at the end of 2000 generations. One can see that populations of smaller networks are less diverse. However, environmental demand (*S*) influences genotypic diversity only weakly, and not in the same unimodal way as seen in Figure S3. Thus, population dynamic processes alone cannot explain the pattern observed in Figures 5A and S3.

The second candidate explanation is that the patterns of Figures 5A and S3 may simply reflect how the number of novel phenotypes in the neighborhood of random metabolic networks varies with *N* and *S*. Figure 5C shows the number of novel phenotypes in the neighborhood of random viable metabolic networks of varying size, and with varying environmental demands on the network. This figure is based on random samples of 200 metabolic networks (see Methods) for every value of *N* and *S* (20 metabolic networks for 10 different sets of sulfur sources at each *S*). The vertical axis of this figure shows the mean and standard deviation of the number of unique novel phenotypes in the neighborhood of the examined networks. It shows similar unimodal characteristics as the data in Figures 5A and S3. The figure demonstrates that the number of novel phenotypes depends strongly on metabolic network size for environments with $S<20$. In this regime, larger metabolic networks have more novel phenotypes in their neighborhood than smaller networks. For $S>20$, the dependency on metabolic network size disappears and the number of accessible novel phenotypes declines again. In sum, the different accessibility of novel phenotypes in evolving populations, at least qualitatively, emerges from how novel phenotypes are distributed in genotype neighborhoods, and how this distribution depends on $S$ and $N$.

## Discussion

The genotype-phenotype map we characterized here shows both similarities and differences to previously characterized such maps [2, 3, 4, 5, 6]. One similarity is the existence of connected genotype networks that extend far through genotype space, and that link genotypes having the same phenotype. Connected sets of metabolic networks viable on the same set of sulfur sources exhibit large maximum genotype distances $D$. For example, networks with as few as 200 reactions can show $D>0.7$, meaning that they share fewer than 30 percent of their reactions. A second similarity regards phenotypic innovations, genotypes whose phenotypes allow viability on novel sulfur sources. The neighborhoods of two genotypes $G_1$ and $G_2$ tend to contain very different phenotypic innovations, even if $G_1$ and $G_2$ are only moderately different. Both features, taken together, facilitate the exploration of novel phenotypes. They would allow a population of organisms (networks) to explore different regions of genotype space, preserving their phenotype while exploring many novel phenotypes.

A major difference to previously studied genotype-phenotype maps regards the relationship between a phenotype's robustness to mutation and a population's ability to explore novel phenotypes. In macromolecules, this relationship appears to be positive: Greater robustness facilitates innovation [9, 10]. Although robust molecules can access, on average, fewer novel phenotypes in their mutational neighborhoods, populations of robust molecules can spread faster through genotype space. In balance, the second process dominates and allows evolving populations to access more novel phenotypes through mutations.

In sulfur metabolism, we do not see this relationship. Robust phenotypes in this context are characterized by viability on few sulfur sources. They are less easily disrupted through eliminations of individual reactions. We found that the number of phenotypic innovations such phenotypes can access in their neighborhood -- through changes of single reactions -- is highest at intermediate robustness, that is, for phenotypes viable on approximately 20 out of 60 carbon sources we examined. (It can also depend on metabolic network size, being lowest for small networks.) This phenomenon cannot solely be explained by the evolutionary dynamics of evolving populations, partly because populations whose members have intermediate robustness do not spread fastest through genotype space. Instead, the phenomenon is a simple consequence of how many novel phenotypes occur in the neighborhoods of individual genotypes. This number peaks for genotypes whose phenotypes have intermediate robustness. It shows the same qualitative dependence on robustness as the number of novel phenotypes accessible to populations. Thus, in this case, population dynamics do not dominate the process of novel phenotype exploration. We note that the total number of possible novel phenotypes decreases exponentially with the number $S$ of sulfur sources on which a network is already viable. If we took this exponential decrease into account, for example by determining the cumulative fraction instead of the number of novel phenotypes accessible to evolving population, this fraction would decrease with increasing $S$.

These observations raise the question whether they are unique to sulfur metabolism or whether they occur in other metabolic systems. As we stated earlier, part of our motivation to study sulfur metabolism was to avoid the much larger number of reactions of carbon metabolism, which render population approaches like ours computationally intractable. Nonetheless, very limited analyses for carbon metabolism are possible. Figure S4A and S4B show the results of such an analysis, based on a small number of populations of networks at moderate size. The analysis has large uncertainties, but it shows a pattern that is at least reminiscent of sulfur metabolism: Innovation peaks at intermediate robustness (the number of alternative carbon sources a phenotype is viable on).

Taken together these analyses show that the organization of different phenotypes in genotype space can differ greatly among different classes of biological systems, such as proteins and metabolic networks. And these differences can affect the ability of a system to explore novel phenotypes in this genotype space.

A third class of analyses regards features that have not been studied previously, partly because they are unique to metabolic systems and our representation of them. One of them regards the analysis of networks with different sizes (numbers of reactions). Our genotype representation can accommodate and allows us to compare networks of different sizes, whereas commonly used representations of other systems -- molecules or regulatory circuits -- cannot. For example, proteins of different length form genotype spaces of different dimensions, making their comparison challenging [18].

When analyzing metabolic networks of different sizes, we found that populations of small networks can explore fewer novel phenotypes (Figures 5A and S3). This observation is easily explained if one considers that populations of such networks have more essential reactions. Their genotype can thus be altered less easily. In consequence, they are genotypically less diverse (Figure 5B), which restricts their access to novel phenotypes (Figure 5B).

Another analysis focusing on network sizes is our characterization of minimal metabolic networks, networks in which all reactions are essential. While the process of genome and metabolic network reduction leading to small networks has been studied for specific biological networks [19], our approach does not start from such a network and can thus provides a more systematic exploration of genotype space. In our analysis of random minimal metabolic network viable on the same sulfur sources, we found that such networks can have large genotype distance. We can explain part of this observation through reactions that are very similar but differ in one of several highly related metabolites. For example, in many types of reactions involving the phosphorylation of a metabolite, the phosphor group donor can be any of ATP, ADP, AMP or even other phosphorylated nucleotide bases. This allows single reactions to be substituted by similar reactions that only use another group donor metabolite. Also, many alternate pathways require only the swapping of two reactions allowing metabolic networks with very few non-essential reactions (little robustness) to substitute some of their reactions. However, these may not be the only explanations of different network architectures, because minimal metabolic networks viable on the same sulfur sources can have dramatic pathway differences (results not shown). Whether such differences can be bridged through series of single reaction changes is a question for future exploration.

Properties of minimal networks also are useful in explaining the maximal genotype distance in a genotype network. For example, for metabolic networks of a given size $N$ and viability on $S$ sulfur sources, the maximum genotype distance within a genotype network is well approximated by one minus the fraction of superessential reactions in minimal metabolic networks. These are reactions found in all minimal networks viable on a given number of sulfur sources. We currently have no mechanistic explanation for this relationship and it, also, remains a subject for future work.

Studies like ours have multiple limitations, including a focus on biomass production phenotypes, limited knowledge of the reaction universe, computational constraints, uncertainty about the most relevant sulfur sources, about thermodynamic factors, about the role cellular compartmentalization, and many others. Although they uncover generic features of genotype-phenotype maps, with demonstrated relevance for evolutionary adaptation and innovation in other biological systems [9, 20], they are just a beginning in the exploration of a vast metabolic genotype space.

# Methods

## Global set of sulfur-involving reactions

To obtain the global set of reactions involving sulfur-containing metabolites that can be present in the metabolic networks we studied, we used data from the LIGAND database of the Kyoto Encyclopedia of Genes and Genomes (KEGG; http://www.genome.ad.jp/kegg/ligand.html) [11]. The LIGAND database is a database of chemical compounds and reactions in biological pathways that was compiled from pathway maps of metabolism of carbohydrates, energy, lipids, nucleotides, amino acids and others. Also included in the database are the list of catalyzed reactions categorized by the Nomenclature Committee of the International Union of Biochemistry and Molecular Biology (NC-IUBMB) (http://www.chem.qmul.ac.uk/iubmb/enzyme/) which includes all enzymes with known classification [21].

Specifically, we used the REACTION and COMPOUND sections of the LIGAND database to construct our global reaction set. From this dataset we pruned (i) all reactions involving general polymer metabolites of unspecified numbers of monomer units ($C_2H_6(CH_2)_n$), or, similarly, general polymerization reactions that were of the form $A_n + B \rightarrow A_{n+1}$, because their abstract form makes them unsuitable for stoichiometric analysis, (ii) reactions involving glycans, because of their complex structure, (iii) reactions that were not stoichiometrically or elementally balanced, and (v) reactions involving complex metabolites without chemical information about their structure.

In addition, we merged all the reactions existing in the *E. coli* metabolic network model (iJR904) [22] that involve sulfur containing compounds. After these steps of pruning and merging, our global reaction set consisted of 1221 reactions.

## Flux balance analysis

Flux balance analysis is a computational method used to find a set of fluxes through all metabolic reactions that maximize biomass production in a given metabolic network, assuming it is in a steady state [12]. This assumption means that the concentrations of internal metabolites does not change over time. To compute the maximum biomass growth using this method, one needs to know the stoichiometric coefficients of each reaction, the chemical environment of the cell (the set of upper bounds on the fluxes of external metabolites into the cell), and the biomass composition, which represents metabolite consumption during cell growth. This consumption is reflected in a "biomass growth reaction", for which we chose the reaction defined in the *E. coli* iJR904 metabolic model [22]. This biomass growth reaction includes all 20 proteinaceous amino acids, nucleotides, deoxynucleotides, putrescine, spermidine, 5-methyltetrahydrofolate, coenzyme-A, acetyl-CoA, succinyl-CoA, cardiolipin, FAD, NAD, NADH, NADP, NADPH, glycogen, lipopolysaccharide, phosphatidylethanolamine, peptidoglycan, phosphatidylglycerol, phosphatidylserine and UDPglucose. For the purpose of this study we concentrated only on the ability of a metabolic network to synthesize the sulfur containing biomass precursors, which are the two amino-acids cysteine and methionine, coenzyme-A, acetyl-CoA and succinyl-CoA. We thus allowed the metabolic networks to uptake any metabolites not containing sulfur. We consider a metabolic network to be *viable* in a given environment if it can sustain a biomass growth rate greater than $1.0\times10^{-3}$. In essence, the approach we take is equivalent to asking whether all the necessary sulfur containing biomass precursors are synthesizable given a metabolic network in a specified environment. Flux balance analysis relies on linear programming [23] to compute the maximum biomass production rate. We used the packages CPLEX (11.0, ILOG; http://www.ilog.com/) and CLP (1.4, Coin-OR; https://projects.coin-or.org/Clp) to solve the associated linear programming problems.

**Environments and phenotypes**

We here considered 124 different environments that differed in the chemical compound that could serve as the *sole* source of sulfur. These 124 sources were all the sulfur containing metabolites in the 1221 reactions of our global reaction set. We provided any metabolite not containing sulfur in the environment, in effect making it a rich environment limited by sulfur containing metabolites only. Also, we allowed cells to secrete all metabolites. We define a metabolic *phenotype* as the set of environments (each with a different sole sulfur source) in which a metabolic network is viable. The *environmental demands* imposed on a metabolic network correspond to the set of sulfur sources that the metabolic network must *at least* be viable in.

**Essential and super-essential reactions**

We define a reaction as essential if its removal from a metabolic network renders the metabolic network inviable on at least one of the sulfur sources that it had previously been viable on. We called a reaction super-essential if it occurred in all minimal metabolic networks generated under a given set of environmental demands.

**Generating random and minimal metabolic networks**

We generated random viable metabolic networks as follows. First, we generated a random environmental demand, that is, we required viability in some given number $X$ of sulfur sources. To this end, we first created a binary vector of length 124 (each of whose entries corresponds to one sulfur source), initialized all its entries to the value zero, and then randomly changed $X$ of these entries to one. These entries represent the set of sulfur sources on which we required our metabolic networks to be viable.

We then generated random viable metabolic network of $N$ reactions as follows. We started from a metabolic network that contained all 1221 reactions (this networks is viable on all 124 sulfur sources) and sequentially removed randomly chosen reactions, while ensuring viability on the set of $X$ sulfur sources chosen previously, until we had reached a network with the target number $N$ of reactions.

We define a minimal metabolic network as a network where not a single reaction can be removed without destroying viability. To generate a (random) minimal metabolic network we used the same procedure until no reactions could be removed without destroying viability.

**Metabolic network random walk maintaining viability in the environmental demands**

We generated random walks for metabolic networks of given reaction numbers N and viability on a given number of sulfur sources by first generating a random metabolic network of this size, as just described. We then generated a series of steps ("mutations") in metabolic genotype space, each one either an addition or a deletion of a reaction. After each step, we recomputed the phenotype of the metabolic network. If the metabolic network was still viable on the same set of sulfur sources, we accepted the mutation and proceeded to the next mutation; if not, we rejected the mutation and repeated the process from the metabolic network prior to the mutation. We continued the resulting random walk for 10,000 accepted mutations. We kept the size of the metabolic network in the narrow interval (N, N+1) by ensuring that accepted mutations alternated between reaction additions and deletions.

In a variation on this procedure, we also carried out forced random walks through genotype space. Their aim was to obtain metabolic networks that are as different (in terms of genotype distance) as possible from the initial metabolic network. In a forced random walk, we required that any reaction addition did not involve a reaction that had been part of the initial network at the start of the walk.

**Population dynamics**

Populations where the product of population size and mutation rate is much greater than one are polymorphic most of the time, and show evolutionary dynamics different from those of small populations [24]. To understand their evolution, one needs to simulate them explicitly. To this end, we implemented a Fisher-Wright model of evolution [25] in populations of 100 metabolic networks. We initialized each population with 100 copies of a single viable metabolic network, and then exposed it to repeated "generations" of mutation (one reaction addition or deletion per network and generation) and selection. Specifically, for the selection procedure, we chose 100 viable individuals at random with replacement to form the next generation. (If a mutation had rendered a network inviable, it could not be chosen.) Our simulations proceeded for 2000 generations.

## Authors' contributions

JFMR and AW designed the study. JFMR carried out all computational analysis. JFMR and AW participated in data analysis and manuscript writing.

## Acknowledgements

We would like to acknowledge support from Swiss National Science Foundation grant 315230_129708, as well as from the YeastX project of SystemsX.ch.

## Figures

Figure 1 – Genotype-phenotype map of metabolic networks. Different representations of a hypothetical metabolic network (A), as a node in a genotype network (B), or as a binary vector (C) listing the reactions in the network. Each genotype (circles) on the genotype network in (B) has 1221 neighbors (not all edges are drawn) that differ by a single mutation. Neighbors in (B) are connected by edges. The colors of the genotypes represent different phenotypes. The phenotypes of the metabolic networks are computed using flux balance analysis applied to 124 environments with different sulfur sources. Two hypothetical phenotypes are represented in (D) as binary vectors listing the environments a genotype is viable in (D). Random evolutionary walks can be seen as paths on a genotype network that stay on genotypes with the same phenotype (represented as the genotype color). "Mutations" correspond to additions or deletions of individual reactions from the metabolic network. The number of genotypes in the genotype space is $2^{1221}$.

Figure 2 - (A) Distribution of genotype distance between pairs of minimal metabolic networks viable under the same environmental demands. (B) Average size (closed circles) and average number of superessential reactions (open circles) of 1000 minimal metabolic networks as a function of environmental demands *S* on a network. The number of superessential reactions was obtained by counting the number of reactions common to 100 minimal metabolic networks generated with the same set of sulfur sources. For each data point, we used 10 different sets of sulfur sources with size *S*. Error bars represent the standard deviations of the distributions.

Figure 3 – Maximum genotype distances and fraction of essential reactions depend strongly on the number of superessential reactions and the sizes of minimal metabolic networks. (A) Average maximum genotype distance for metabolic networks of different sizes and subject to different environmental demands after random walks of 10'000 accepted reaction changes. (B) Fraction of essential reactions found in random metabolic networks of different size and subject to different environmental demands. Each data point is an average over 200 random

walks (20 random walks for 10 different sets of environmental demands with the same number *S* of sulfur sources).

Figure 4 – Fraction of phenotypes unique to the neighborhood of an evolving metabolic network $G_k$ (vertical axis) when compared to the neighborhood of a starting metabolic network *G* after *k* "mutations", i.e., reaction changes (horizontal axis). Each data point represents an average over 200 evolving metabolic networks of size *N*=200 (20 random walks for 10 different sets of environmental demands with the same number *S* of required sulfur sources).

Figure 5 - Metabolic networks have diverse phenotypes in their neighborhood. (A) Cumulative number of novel phenotypes encountered in the neighborhoods of evolving metabolic networks of different sizes and subject to different environmental demands. (B) The average pairwise genotype distance found in populations of evolving metabolic networks. Each population consists of 100 individual metabolic networks. (C) Number of novel phenotypes found in the (1-mutant) neighborhood of random metabolic networks of different size *N* and subject to different environmental demands *S*. Each data point is an average over 200 metabolic networks (20 random walks for 10 different sets of environmental demands, with the same number *S* of required sulfur sources).

## Additional files

Figure S1 – Average number of sulfur sources that random metabolic networks are actually viable in, for varying environmental demands *S*, and varying metabolic network size *N*. The figure demonstrates that random metabolic networks required to be viable on a given number *S* of sulfur sources (as generated by the procedures described in Methods) are generally viable on more than *S* sulfur sources. Each data point represents an average over 200 random metabolic networks (20 random metabolic networks generated under 10 different sets of environmental demands with the same number *S* of required sulfur sources). Error bars correspond to one standard deviation.

Figure S2 – Number of essential reactions found in random metabolic networks of different size and for different environmental demands (*S*). Each data point represents an average over 200 random metabolic networks (20 random metabolic networks generated under 10 different sets of environmental demands with the same number *S* of required sulfur sources).

Figure S3 – Cumulative number of novel phenotypes encountered in the neighborhoods of all evolving metabolic networks in a large population. The results are plotted for populations of metabolic networks of different sizes and subject to different environmental demands. Each data point represents an average over 200 simulations, 20 simulations for 10 different sets of environmental demands with the same number *S* of sulfur sources. Each population consisted of 100 individual metabolic networks.

Figure S4 – Plot of the cumulative number of novel phenotypes found in the neighborhood of (A) large and (B) small evolving populations of metabolic networks required to be viable in different number of carbon sources. (C) Number of novel carbon utilization phenotypes found in the neighborhood of random metabolic networks. Metabolic networks in these simulations had 931 reactions, the same as the size of the *E. coli* iJR904 model [5, 22].

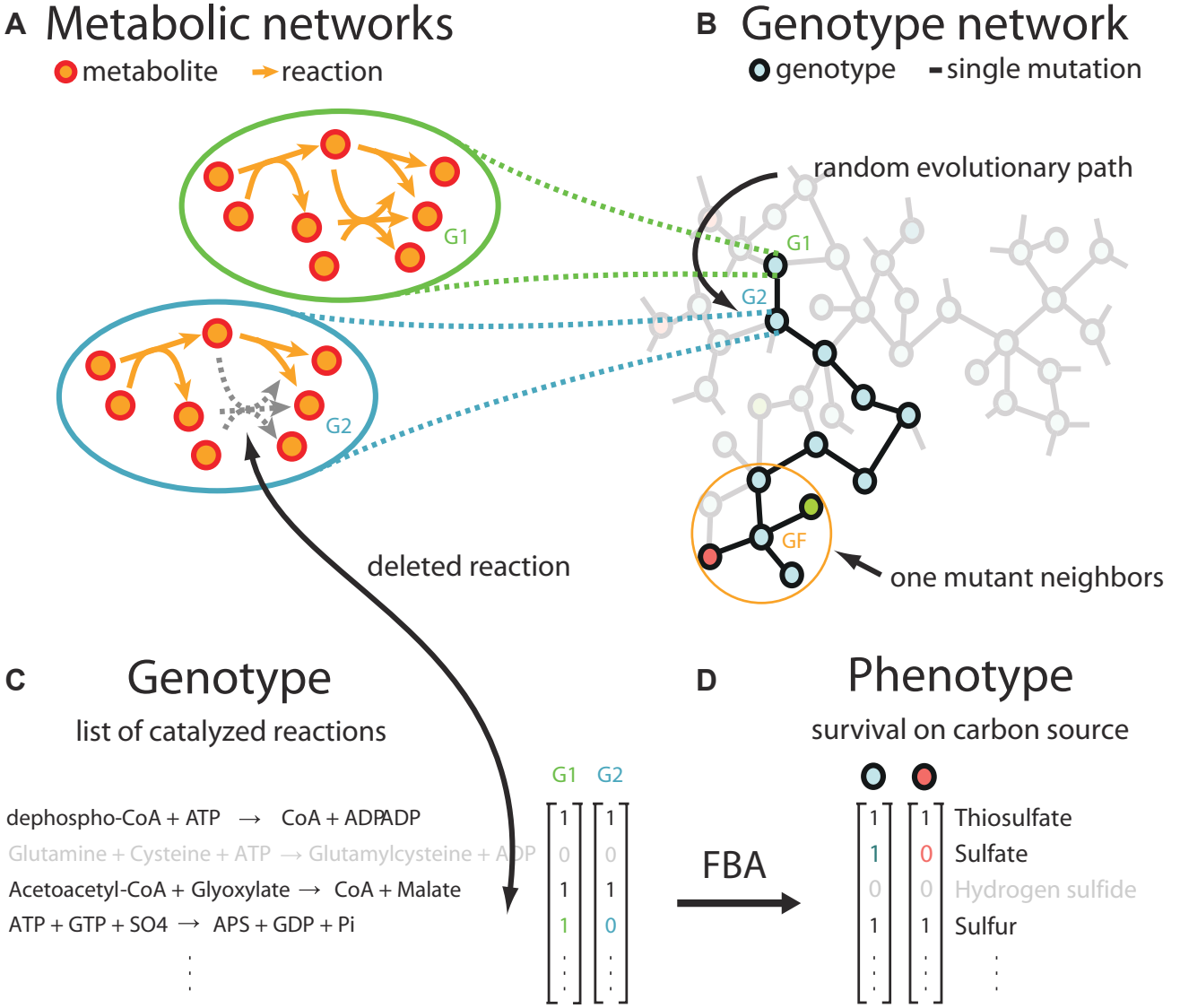
A Metabolic networks
metabolite
reaction
G1
G2
B Genotype network
genotype
single mutation
random evolutionary path
G1
G2
GF
one mutant neighbors
deleted reaction
C Genotype
list of catalyzed reactions
dephospho-CoA + ATP → CoA + ADPADP
Glutamine + Cysteine + ATP → Glutamylcysteine + ADP
Acetoacetyl-CoA + Glyoxylate → CoA + Malate
ATP + GTP + SO4 → APS + GDP + Pi
G1 G2
1 1
0 0
1 1
1 0
FBA
D Phenotype
survival on carbon source
1 1 Thiosulfate
1 0 Sulfate
0 0 Hydrogen sulfide
1 1 Sulfur

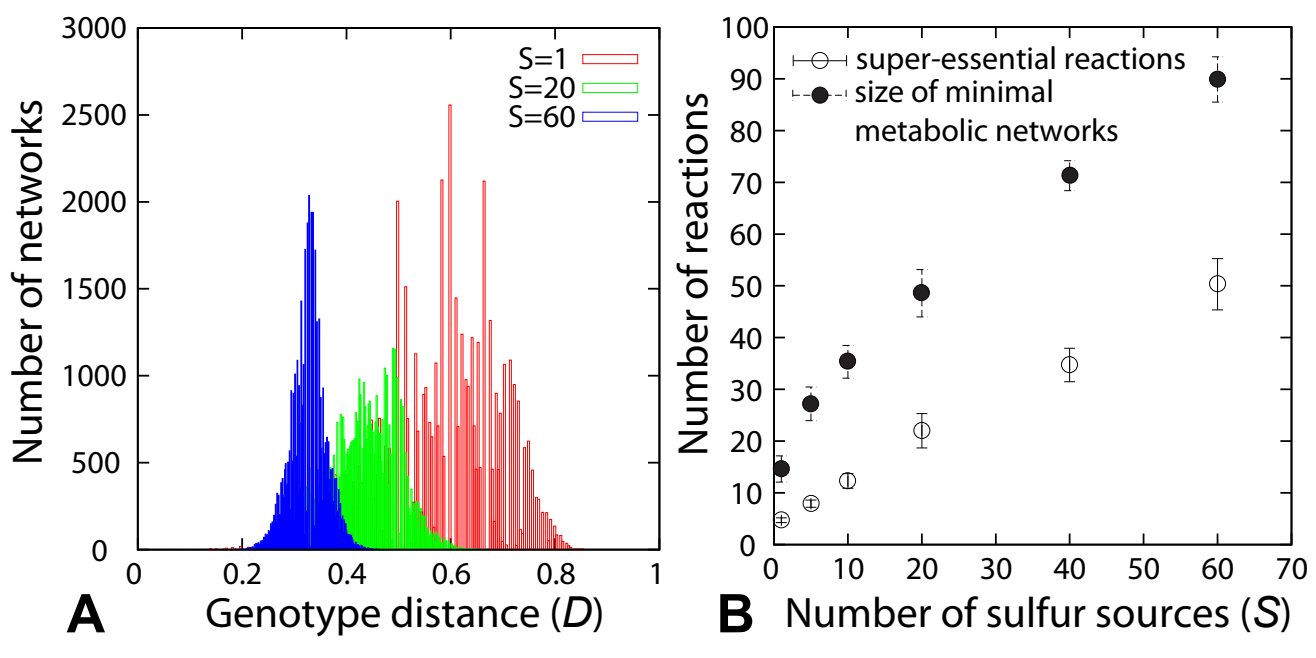

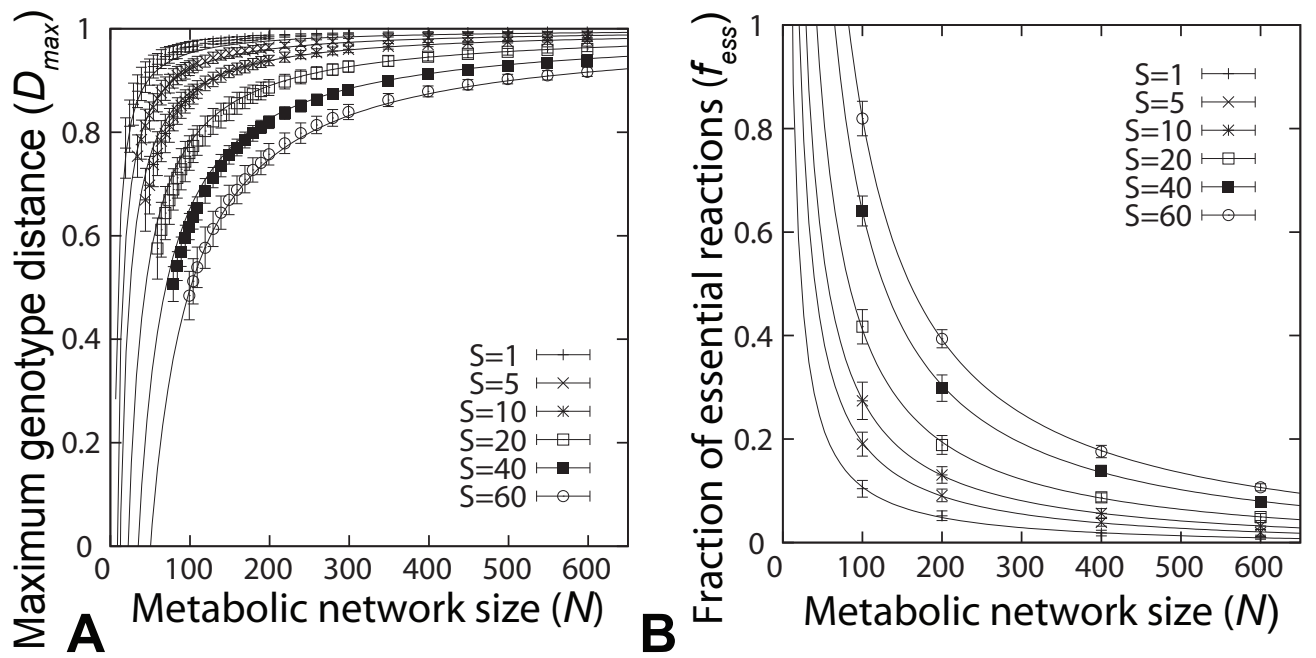

Maximum genotype distance ($D_{max}$)
Metabolic network size ($N$)
S=1
S=5
S=10
S=20
S=40
S=60
A
Fraction of essential reactions ($f_{ess}$)
Metabolic network size ($N$)
S=1
S=5
S=10
S=20
S=40
S=60
B

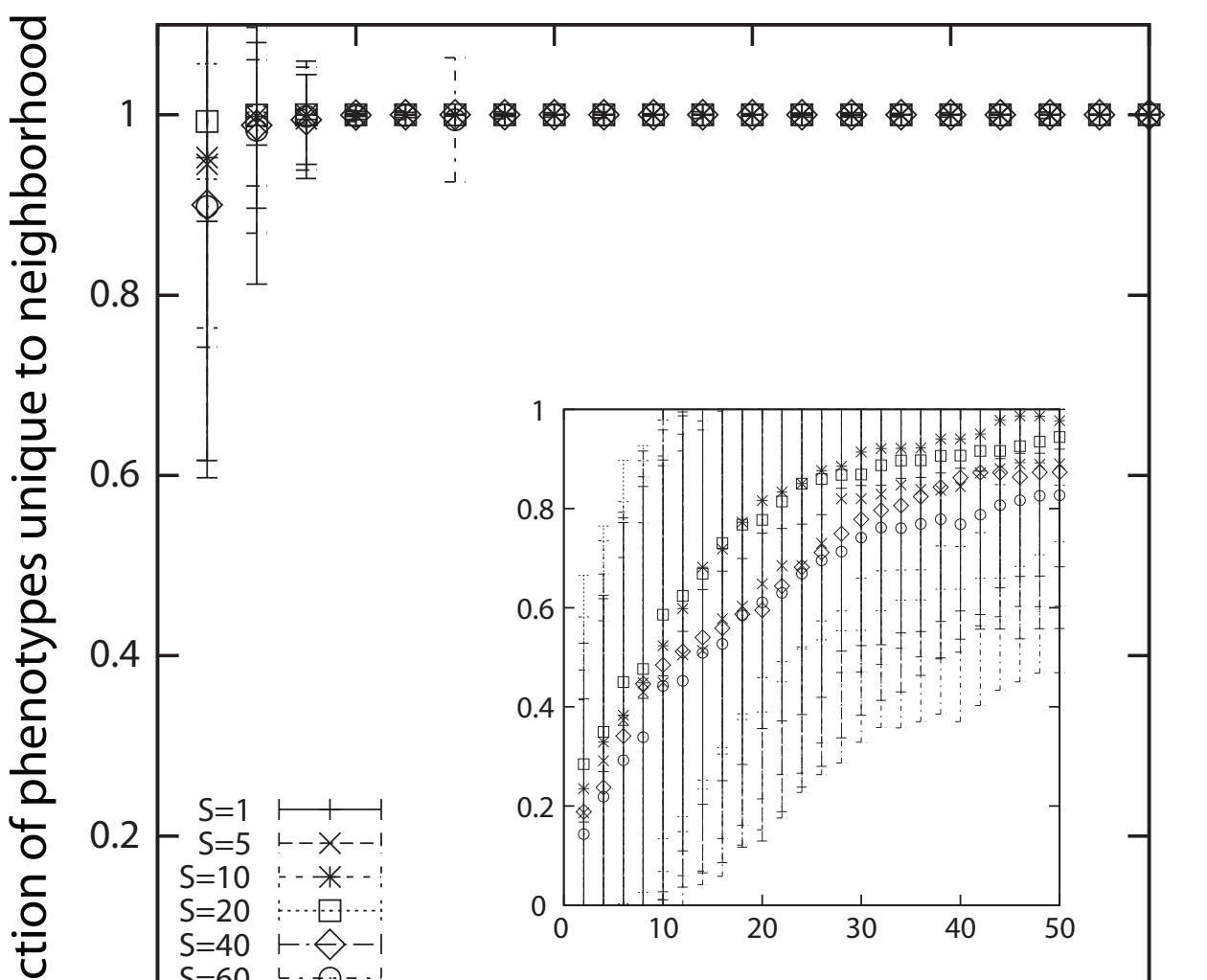

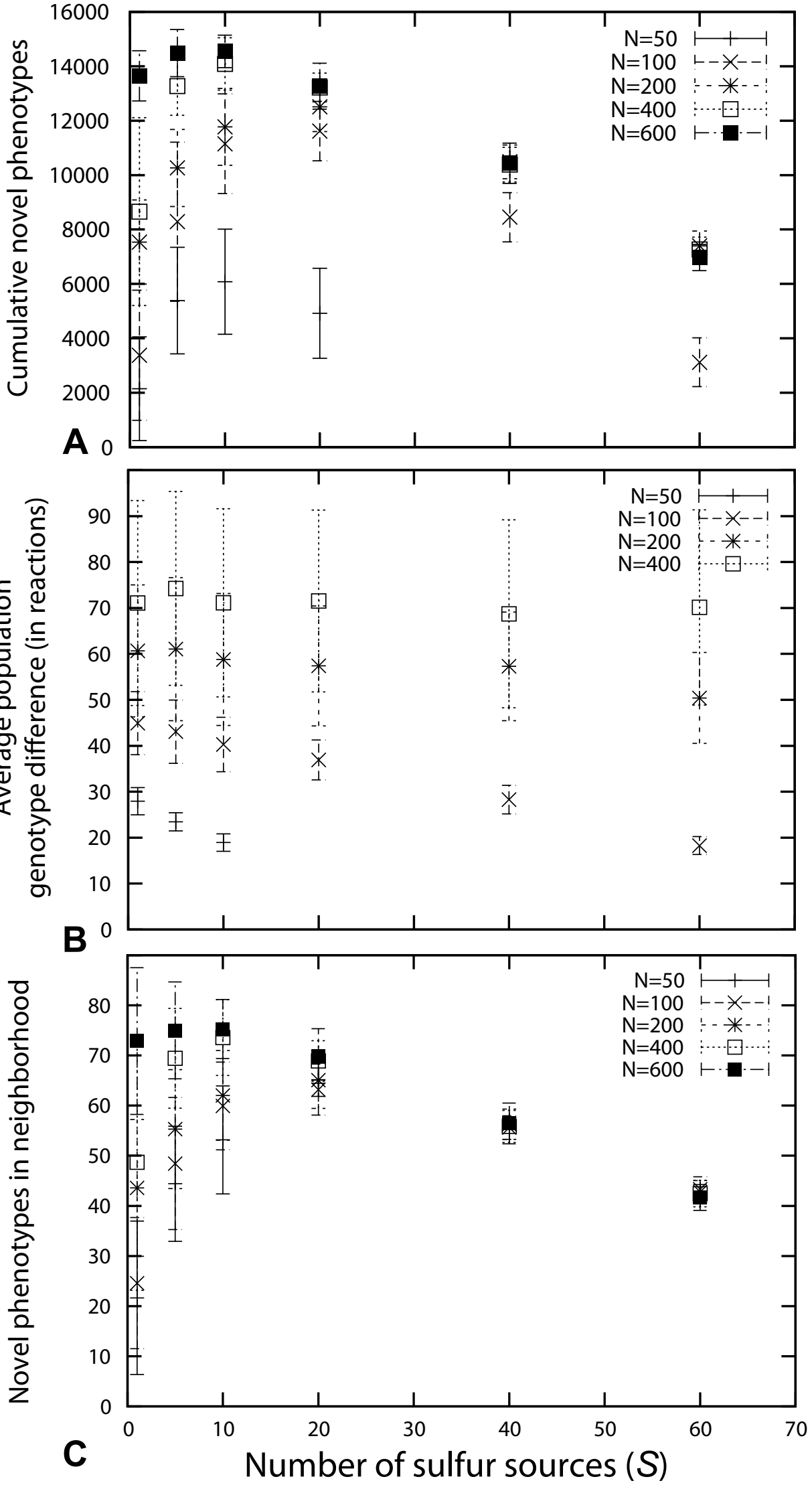

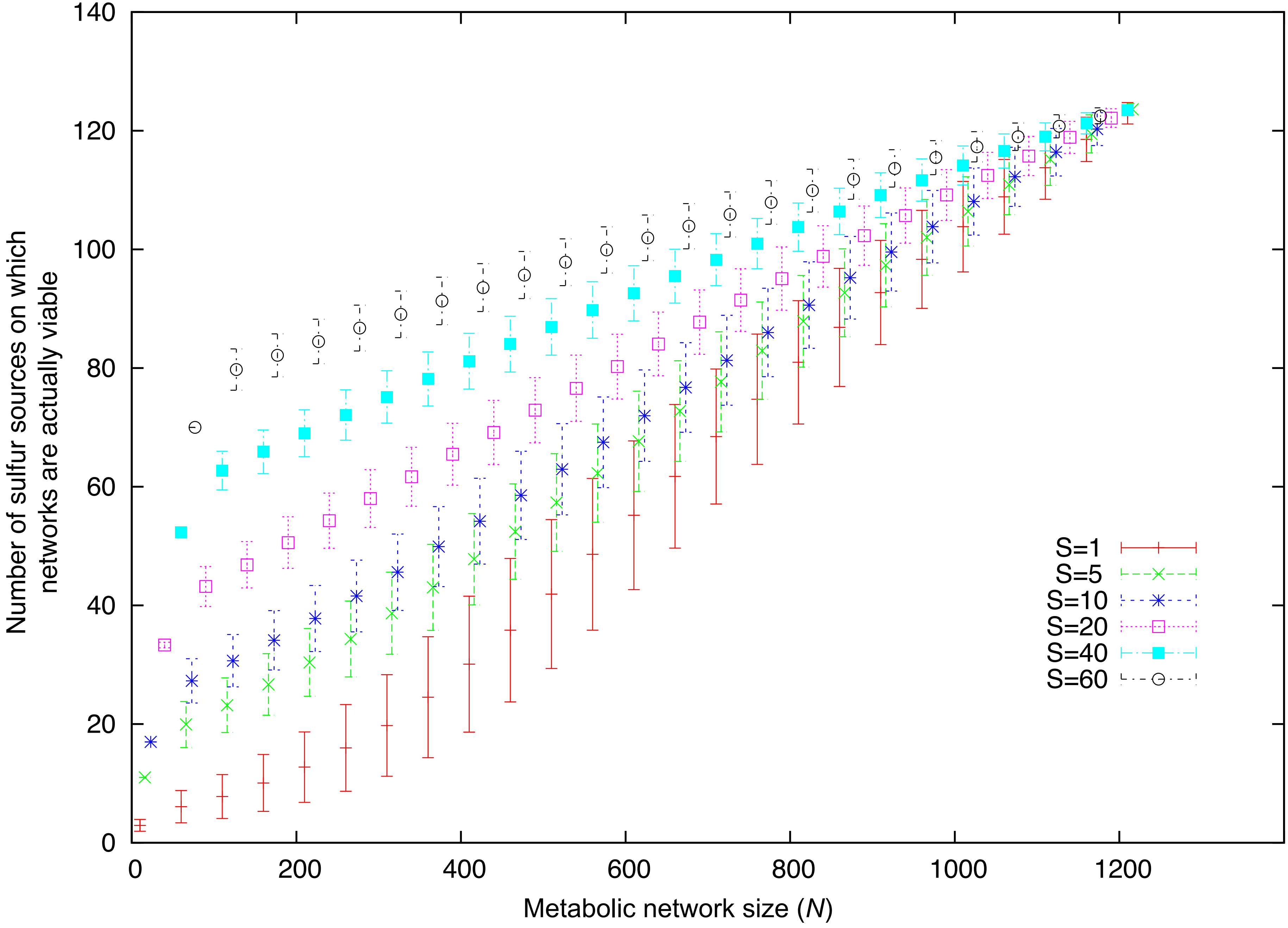

Number of sulfur sources on which networks are actually viable
Metabolic network size (N)
0
20
40
60
80
100
120
140
0
200
400
600
800
1000
1200
S=1
S=5
S=10
S=20
S=40
S=60

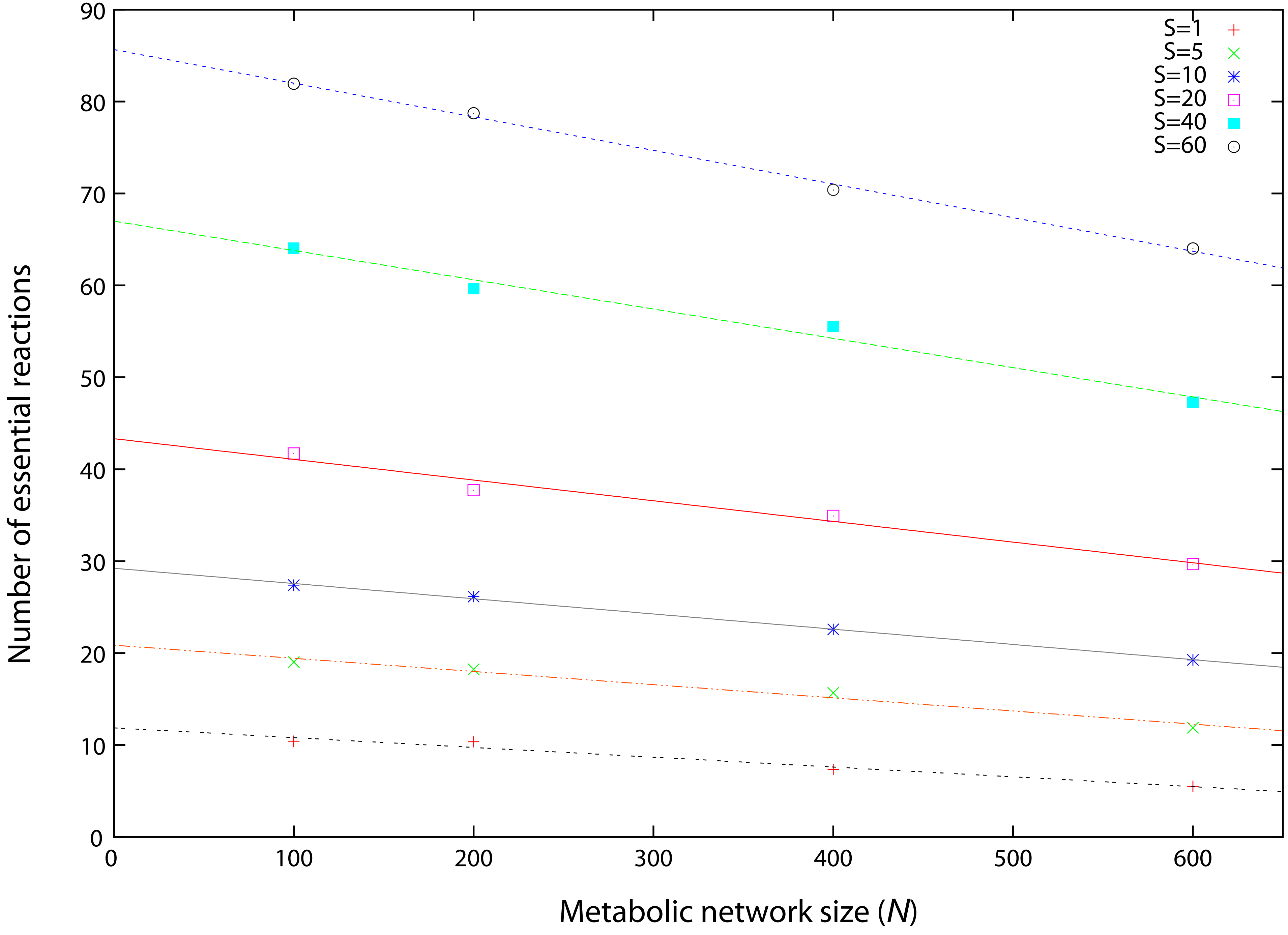

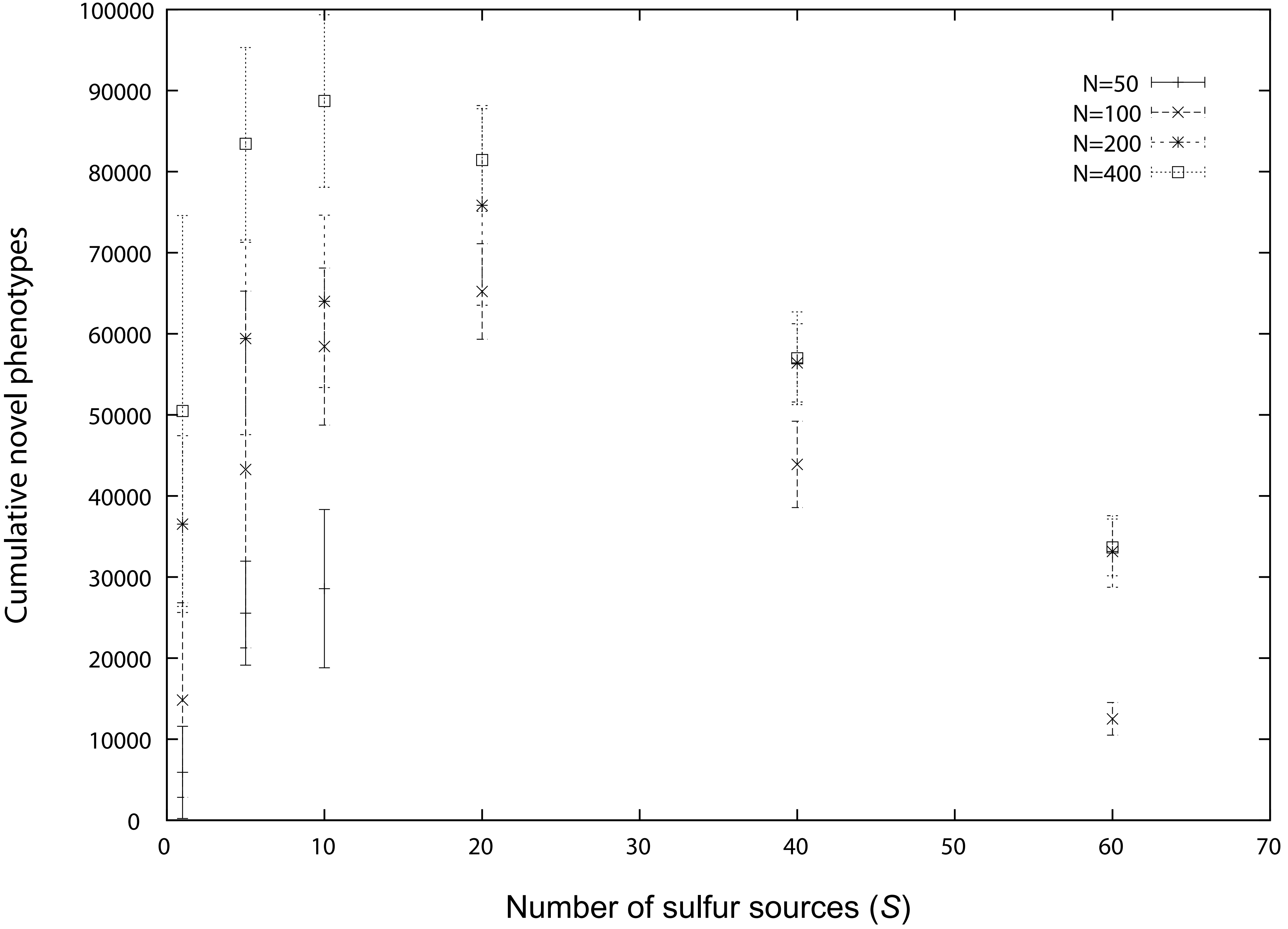

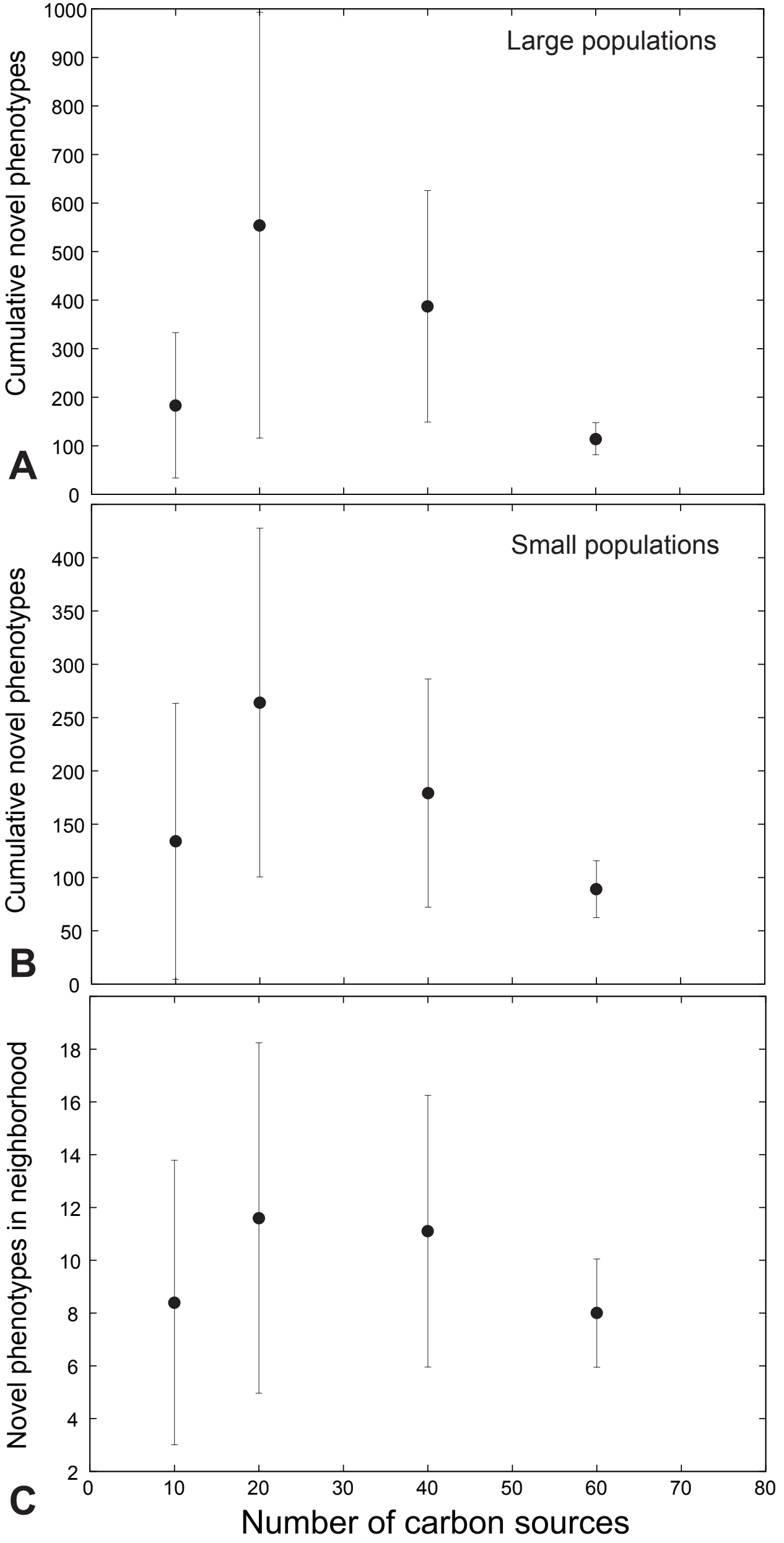